\documentclass[11pt]{article}
\usepackage[dvips]{graphicx}
\usepackage{mathrsfs}
\usepackage{amsthm} 
\usepackage{amsmath}
\usepackage{amssymb}
\usepackage{amsfonts}
\usepackage{float}
\usepackage{yhmath}
\usepackage{wrapfig}
\usepackage[dvips]{color}
\usepackage[dvips]{colortbl}
\usepackage{cite}
\usepackage{array}
\usepackage{tabularx}
\usepackage[sectionbib]{chapterbib}
\usepackage{threeparttable}
\usepackage{chemarrow}
\usepackage{framed}
\usepackage{ascmac}
\definecolor{shadecolor}{gray}{0.90}
\usepackage[small, bf, labelsep=quad]{caption}
\usepackage{cases}

\newcommand{\mb}[1]{\textnormal{\mathversion{bold}{$#1$}}}

\newcommand{\G}{\textit{\textsf{G}}}
\newcommand{\HG}{\hspace{1.3mm}\Hat{\textit{\textsf{\hspace{-1.3mm}G}}}\hspace{0.5mm}}
\newcommand{\I}{I}
\newcommand{\II}{I\hspace{-0.5mm}I}

\definecolor{mycolor1}{rgb}{1,1,0.7}
\definecolor{mycolor2}{rgb}{0.9,1,1}
\definecolor{mycolor3}{cmyk}{0,0,0,0.113}
\definecolor{mycolor4}{cmyk}{0.086,0,0,0}

\begin{document}

\begin{center}
\Huge{Critical Nature of the Size Exponent of Polymers}
\end{center}

\vspace*{5mm}
\begin{center}
\large{Kazumi Suematsu\footnote{\, The author takes full responsibility for this article.}, Haruo Ogura$^{2}$, Seiichi Inayama$^{3}$, and Toshihiko Okamoto$^{4}$} \vspace*{2mm}\\
\normalsize{\setlength{\baselineskip}{12pt} 
$^{1}$ Institute of Mathematical Science\\
Ohkadai 2-31-9, Yokkaichi, Mie 512-1216, JAPAN\\
E-Mail: suematsu@m3.cty-net.ne.jp, ksuematsu@icloud.com  Tel/Fax: +81 (0) 593 26 8052}\\[3mm]
$^{2}$ Kitasato University,\,\, $^{3}$ Keio University,\,\, $^{4}$ Tokyo University\\[10mm]
\end{center}

\hrule
\vspace{3mm}
\noindent
\textbf{\large Abstract}: On the basis of the thermodynamic theory of the excluded volume effects, we show that the size exponent varies abruptly, depending on the change of the segment concentration. For linear polymers, the exponent changes discontinuously from $\nu=3/5$ for the isolated system ($\bar{\phi}=0$) in good solvents to $\nu=1/2$ in the finite concentration ($0<\bar{\phi}\le1$), while for branched polymers having $\nu_{0}=1/4$, the corresponding exponent varies from $\nu=1/2$ ($\bar{\phi}=0$) to $\nu\cong 1/3$ ($0<\bar{\phi}\le1$). 

\vspace{0mm}
\begin{flushleft}
\normalsize{}\\[3mm]
\end{flushleft}
\hrule
\vspace{10mm}
\setlength{\baselineskip}{14pt}

\vspace{0mm}
\section{Introduction}
Making demarcation of polymer solutions according to concentration ranges (isolated, dilute, semi-dilute, medium, concentrated, and melt) is often useful since polymer solutions manifest different facets depending on the respective polymer concentrations: viscosity, elastic behavior, scattering effects, excluded volume effects, and so forth. Such classification, on the other hand, fails as far as the evaluation of the exponent, $\nu$, is concerned. This is because the exponent is defined for the thermodynamic limit: $N\rightarrow\infty$. It is on the invalidity of such classification in the determination of the exponent that we are going to discuss in this paper.

To discuss the above issue rigorously, we must return to the starting point of the formulation of the Gaussian approximation for the expanded polymers.

\section{Problem on Theoretical Legitimacy and Presentation of a Proper Formulation}
As mentioned in the previous work\cite{Kazumi}, two approximate expressions for expanded polymers have been used conveniently. One is the following approximation, originally introduced by Flory in the calculation of the expansion factor, $\alpha$\cite{Flory}:
\begin{equation}
p(x,y,z)dxdydz=\left(\frac{\beta}{\pi\alpha^{2}}\right)^{3/2}\exp\left\{-\beta\left[(x-a)^{2}+(y-b)^{2}+(z-c)^{2}\right]\right\}\alpha^{3}dxdydz\label{Nenk-1}
\end{equation}
where $\beta=3/2\langle s_{N}^{2}\rangle_{0}$. Let $\delta V=\alpha^{3}dxdydz$. The segment concentration at the coordinates $(x, y, z)$ is then
\begin{align}
C(x, y, z)&=N\left(\frac{\beta}{\pi\alpha^{2}}\right)^{3/2}\sum_{\{a, b, c\}}\exp\left\{-\beta\left[(x-a)^{2}+(y-b)^{2}+(z-c)^{2}\right]\right\}\notag\\
&=N\left(\frac{\beta}{\pi\alpha^{2}}\right)^{3/2}\G(x, y, z)\label{Nenk-2}
\end{align}
Unfortunately, the above approximation is clearly incorrect since Eq. (\ref{Nenk-1}) is merely the expression for the unperturbed polymer. Moreover, Eq. (\ref{Nenk-2}) is not normalized for the integration over the whole space. To improve on the formulation of Eq. (\ref{Nenk-1}), we put forth the mathematically legitimate approximation\cite{Flory, Kazumi}:
\begin{equation}
\Hat{p}(x,y,z)dxdydz=\left(\frac{\beta}{\pi\alpha^{2}}\right)^{3/2}\exp\left\{-\frac{\beta}{\alpha^{2}}\left[(x-a)^{2}+(y-b)^{2}+(z-c)^{2}\right]\right\}dxdydz\label{Nenk-3}
\end{equation}
Since $\delta V=dxdydz$, Eq. (\ref{Nenk-3}) gives the alternative concentration formula:
\begin{align}
\Hat{C}(x, y, z)&=N\left(\frac{\beta}{\pi\alpha^{2}}\right)^{3/2}\sum_{\{a, b, c\}}\exp\left\{-\frac{\beta}{\alpha^{2}}\left[(x-a)^{2}+(y-b)^{2}+(z-c)^{2}\right]\right\}\notag\\
&=N\left(\frac{\beta}{\pi\alpha^{2}}\right)^{3/2}\HG(x, y, z)\label{Nenk-4}
\end{align}
which is normalized for the whole space. Unexpectedly, when applied to the isolated polymer solution, Eqs. (\ref{Nenk-2}) and (\ref{Nenk-4}) yield the same classic result, the celebrated equality by Flory:
\begin{equation}
\alpha^{d+2}-\alpha^{d}=N^{2}\frac{V_{2}^{\,2}}{V_{1}}\left(\frac{\beta}{\pi}\right)^{\frac{d}{2}}\left\{\frac{1}{2^{\frac{d}{2}}}\left(1/2-\chi\right)+\frac{V_{2}N}{3^{\frac{d}{2}+1}\alpha^{d}}\left(\frac{\beta}{\pi}\right)^{\frac{d}{2}}+\frac{\left(V_{2}N\right)^{2}}{4^{\frac{d}{2}+1}\alpha^{2d}}\left(\frac{\beta}{\pi}\right)^{d}+\cdots\right\}\label{Nenk-5}
\end{equation}
It follows that, as far as we deal with the isolated polymer system, no difference arises between the two formulations (\ref{Nenk-1}) and (\ref{Nenk-3}) (designated $p$ and $\Hat{p}$ respectively)\cite{Kazumi}.

On the other hand, when we apply the above density equations (\ref{Nenk-2}) and (\ref{Nenk-4}) to concentrated systems, we are faced with crucial differences. The key quantity in concentrated solutions is to evaluate the gradient of the Gibbs potential around a polymer. This can be accomplished by taking the difference of $G_{mixing}$ between a more concentrated region $\II$\, and a more dilute region $\I$. The basic thermodynamic equation is
\begin{equation}
\frac{d G}{d\alpha}=\frac{d}{d\alpha}\left(G_{\II}-G_{\I}\right)+\frac{d G_{\text{elasticity}}}{d\alpha}=0\label{Nenk-6}
\end{equation}
which can be recast in terms of the chemical potentials:
\begin{equation}
\frac{d G}{d\alpha}=\left(\mu_{c_{2\II}}-\mu_{c_{2I}}\right)\frac{d c_{2}}{d\alpha}+\frac{dG_{\text{elasticity}}}{d\alpha}=0\label{Nenk-7}
\end{equation}
where $G$ is the Gibbs potential in the system and $\mu$ the chemical potential; $c_{2}$ denotes the concentration of polymer segments. $\mu$ is the quantity associated with the mixing free energy: 
\begin{equation}
\mu_{c_{2}}=\left(\frac{\partial G_{mixing}}{\partial c_{2}}\right)_{T, P}\label{Nenk-8}
\end{equation}
with 
\begin{equation}
G_{mixing}=\,(kT/V_{1})\int (1-v_{2})\left\{\log\,\left(1-v_{2}\right)+\chi v_{2}\right\}\delta V\label{Nenk-9}
\end{equation}
where $V_{1}$ denotes the volume of a solvent molecule, $\chi$ the enthalpy parameter, and $v_{2}$ the volume fraction of segments in the system.
By solving Eq. (\ref{Nenk-7}), the expansion factor, $\alpha$, can be obtained as a function of segment concentrations and molecular weights. It is noteworthy that Eq. (\ref{Nenk-7}) does not include the role of solvents explicitly, showing that the excluded volume effects are phenomena that can be explained by the movement of polymer molecules alone.

Applying Eqs. (\ref{Nenk-1}) and (\ref{Nenk-3}) to the basic formula (\ref{Nenk-6}) or (\ref{Nenk-7}), and converting the notations as $\II\rightarrow hill$ and $\I\rightarrow valley$ in accordance with the notations of polymer physics, we have the following solutions respectively:
\begin{equation}
\alpha^{5} - \alpha^{3}=\left[N^{2}\frac{V_{2}^{\,2}}{V_{1}}\left(1/2-\chi\right)\left(\frac{\beta}{2\pi}\right)^{3/2}\right]\left[\left(\frac{2\beta}{\pi}\right)^{3/2}\iiint\left(\G\hspace{0.3mm}_{hill}^{\,2}-\G\hspace{0.3mm}_{valley}^{\,2}\right)dxdydz\right]\label{Nenk-10}
\end{equation}
for Eq. (\ref{Nenk-1})\footnote{\, We can not apply Eq. (\ref{Nenk-1}) to the formula (\ref{Nenk-7}), since Eq. (\ref{Nenk-7}) combined with Eq. (\ref{Nenk-1}) does not yield an elementary function of $C$ because of the form of $dV=\alpha^{3}dxdydz$. The sound mathematical expression compatible with Eq. (\ref{Nenk-7}) is Eq. (\ref{Nenk-3}), which leads to Eq. (\ref{Nenk-11}). Eq. (\ref{Nenk-1}), on the other hand, can usefully be combined with Eq. (\ref{Nenk-6}) to yield Eq. (\ref{Nenk-10}).}, and
\begin{equation}
\alpha-1/\alpha =-\frac{V_{2}}{3V_{1}}\iiint\left\{\left(1-2\chi\right)\Hat{\mathscr{J}}^{1}+\frac{1}{2}\Hat{\mathscr{J}}^{2}+\cdots\right\}\left(\frac{\partial\Hat{C}}{\partial \alpha}\right) dx dy dz\label{Nenk-11}
\end{equation}
or the mathematical equivalent:
\begin{equation}
\alpha-1/\alpha =-\frac{1}{3V_{1}}\frac{\partial}{\partial\alpha}\iiint\left\{\left(1/2-\chi\right)\Hat{\mathscr{J}}^{2}+\frac{1}{6}\Hat{\mathscr{J}}^{3}+\cdots\right\}dxdydz\tag{\ref{Nenk-11}$'$}
\end{equation}
for Eq. (\ref{Nenk-3}), where $\Hat{\mathscr{J}}^{k}=\Hat{v}_{2, hill}^{k}-\Hat{v}_{2, valley}^{k}$. In Eqs. (\ref{Nenk-11}) and (\ref{Nenk-11}$'$), we have used the relation: $$\Hat{v}_{2}=V_{2}\Hat{C}=V_{2}N\left(\frac{\beta}{\pi\alpha^{2}}\right)^{3/2}\HG(x, y, z)$$ where $V_{2}$ denotes the volume of a segment (equal to the repeating unit of a polymer), and
\begin{equation}
\HG(x, y, z)=\sum_{\{a, b, c\}}\exp\left\{-\frac{\beta}{\alpha^{2}}\left[(x-a)^{2}+(y-b)^{2}+(z-c)^{2}\right]\right\}\label{Nenk-12}
\end{equation}
as defined above.
The differentiation of the integrand in Eq. (\ref{Nenk-11}$'$) with respect to $\alpha$ immediately recovers Eq. (\ref{Nenk-11}), giving a proof of the soundness of Eq. (\ref{Nenk-11}$'$).

Whereas Eq. (\ref{Nenk-1}) leads to the neat solution represented by Eq. (\ref{Nenk-10}), Eq. (\ref{Nenk-3}) leads us to the complicated mathematics (\ref{Nenk-11}$'$) appearing irreducible anymore. One of the main themes of the present work is to clarify the pros and cons of the approximate solution (\ref{Nenk-10}) in light of the comparison with the legitimate solution  (\ref{Nenk-11}$'$).\\

Before proceeding with the central themes, let us discuss the behavior of linear polymers in finite concentration ranges, which appears not to have fully been discussed in the community.

\section{The Behavior of Linear Polymers in Finite Concentration Ranges}\label{Theo}
The average segment concentration is expressed as
\begin{equation}
\Bar{C}=\frac{nN}{V} \hspace{5mm}\text{or} \hspace{5mm}\Bar{\phi}=V_{2}\frac{nN}{V}\label{Nenk-13}
\end{equation}
where $n$ denotes the number of polymers, $N$ the degree of polymerization, $V_{2}$ the volume of a segment, and $V$ the system volume. Suppose a situation that one wants to deduce the exponent, $\nu$, of a polymer in the semi-dilute regime. One might develop a theory, according to the definition of the exponent, $\left\langle s_{N}^{2}\right\rangle\propto N^{2\nu}$ for $N\rightarrow\infty$, as follows:\\[-2mm]

``\textit{Since the average concentration is defined as above, if $N$ is augmented to infinity while $\Bar{C}$ is retained constant, the quantity $n/V$ must be decreased accordingly. This means that any polymer solutions having a finite concentration, $0<\Bar{\phi}<1$, should approach the dilution limit, $n/V\rightarrow 0$, as $N\rightarrow\infty$. As a consequence, one might be led to the hypothesis that any polymers should have the same exponent, $\nu=\nu_{dilution\,limit}$, over the entire range of the finite segment concentration.}''\\[-2mm]

The above seemingly straightforward and reasonable deduction, however, contradicts violently our observations on linear polymers\cite{Vos, Olaj, Wall, Kazumi}. On the contrary, it is known that, as $N$ increases, the excluded volume effects tend to disappear in all finite concentration ranges, and the exponent approaches $\nu=\nu_{0}$\cite{Kazumi}, the value corresponding to the unperturbed configuration. To see this, it is useful to make a thorough scrutinization of the extended excluded volume theory that was derived by using Eq. (\ref{Nenk-1}):
\begin{equation}
\alpha^{5} - \alpha^{3}=\left[N^{2}\frac{V_{2}^{\,2}}{V_{1}}\left(1/2-\chi\right)\left(\frac{\beta}{2\pi}\right)^{3/2}\right]\left[\left(\frac{2\beta}{\pi}\right)^{3/2}\iiint\left(\G\hspace{0.3mm}_{hill}^{\,2}-\G\hspace{0.3mm}_{valley}^{\,2}\right)dxdydz\right]\tag{\ref{Nenk-10}}
\end{equation}
where 
\begin{equation}
\G=\sum_{\{a, b, c\}}\exp\left\{-\beta\left[(x-a)^{2}+(y-b)^{2}+(z-c)^{2}\right]\right\}\label{Nenk-14}
\end{equation}
is a quantity associated with the segment concentration at the coordinates $(x, y, z)$. To evaluate the inhomogeneity term, we make use of the lattice model\cite{Kazumi}. Each polymer is put on the site on the simple cubic lattice with the unit lengths $p\times p\times p$. For the purpose of numerical simulation, we define the intervals as $[-p/4, p/4]$ for $C_{hill}$, and $[p/4, 3p/4]$ for $C_{valley}$. Hence, the integral in the inhomogeneity term is approximated as
\begin{equation}
J_{\alpha}=\iiint_{-p/4}^{p/4}\G\hspace{0.3mm}^{\,2}\,dxdydz-\iiint_{p/4}^{3p/4}\G\hspace{0.3mm}^{\,2}\,dxdydz\label{Nenk-15}
\end{equation}
Note that unlike $\HG$, $\G$ is independent of $\alpha$, which poses a problem as will be seen shortly (Section \ref{AECP}). In Eq. (\ref{Nenk-10}), the first term represents the classic excluded volume effects (the fifth power rule) and the second term represents the wild inhomogeneity in the solution. In the isolated system, $\G\hspace{0.3mm}_{hill}\rightarrow\exp\left\{-\beta\left(x^{2}+y^{2}+z^{2}\right)\right\}$ and $\G\hspace{0.3mm}_{valley}\rightarrow 0$, so that the second term $\left(2\beta/\pi\right)^{3/2}J_\alpha$ equals unity. It turns out that the inhomogeneity term varies from 1 to zero, acting like a statistical weight.

\begin{figure}[h]
\begin{center}
\begin{minipage}[t]{0.46\textwidth}
\begin{center}
\includegraphics[width=7.3cm]{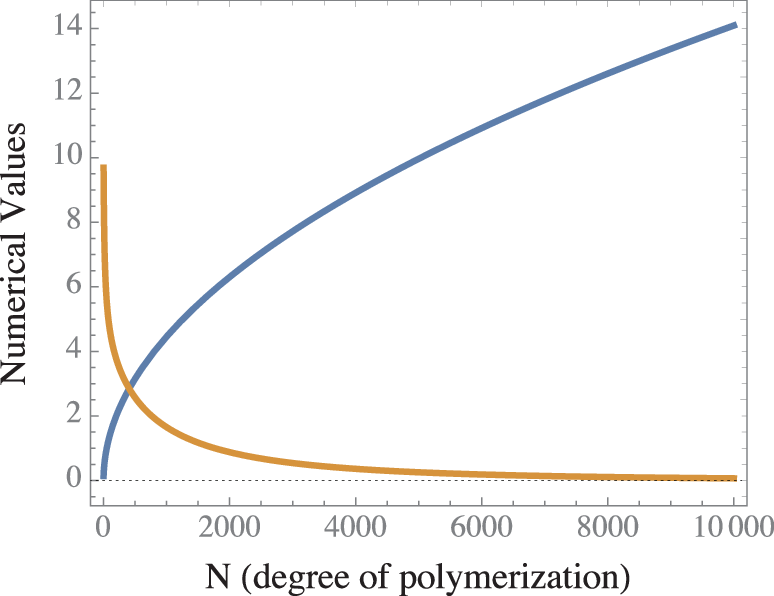}
\end{center}
\vspace{-2mm}
\caption{The variations of the excluded volume (blue line) and the inhomogeneity (red line: magnified ten times) terms as a function of $N$: PSt$-\text{CS}_{2}$ system ($\bar{\phi}=0.1$, $\chi=0.4$).}\label{Nenkai-1}
\end{minipage}
\hspace{10mm}
\begin{minipage}[t]{0.46\textwidth}
\begin{center}
\includegraphics[width=7.3cm]{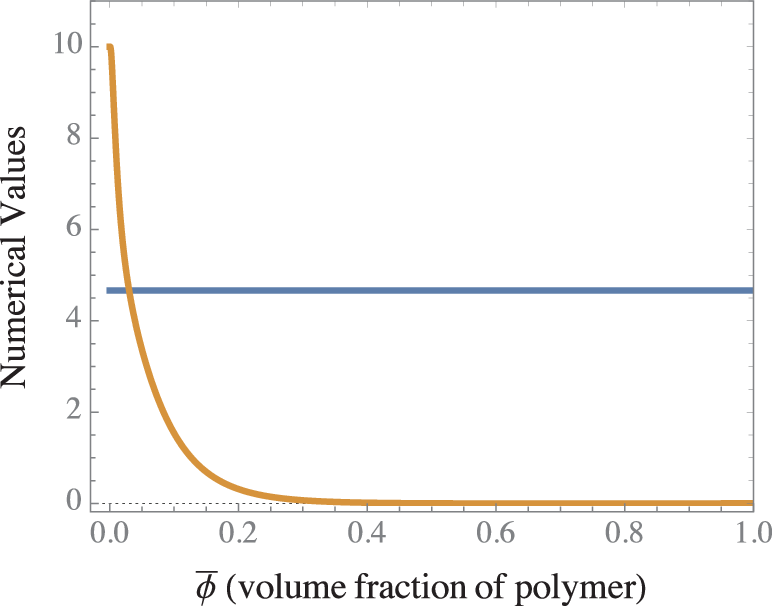}
\end{center}
\vspace{-2mm}
\caption{The variations of the excluded volume (blue line) and the inhomogeneity (red line: magnified ten times) terms as a function of $\Bar{\phi}$: PSt$-\text{CS}_{2}$ system ($N=1100$, $\chi=0.4$).}\label{Nenkai-2}
\end{minipage}
\end{center}
\vspace*{0mm}
\end{figure}

With the help of Eq. (\ref{Nenk-15}), we can show that, as $N$ goes to infinity, (i) whereas the first term increases as $\propto N^{1/2}$, (ii) the second term drops strongly, canceling the increase of the first term. As a result, the product declines rapidly to zero, which leads to $\alpha\rightarrow 1$. This aspect is illustrated in Fig. \ref{Nenkai-1}, which shows the variations of the two terms as a function of $N$ at $\Bar{\phi}=0.1$. The simulation was performed by modeling the PSt$-\text{CS}_{2}$ system ($\chi=0.4$). The blue solid line shows the first term, and the red solid line the second term (magnified ten times). The major point is that, with increasing $N$, the decrease of the inhomogeneity term overwhelms the increase of the first term, and the product $[\textsf{first term}]\times[\textsf{second term}]$ goes to 0. Fig. \ref{Nenkai-2} shows the variation of the second term as a function of $\Bar{\phi}$ with $N=1100$ being fixed. It is seen that with increasing $\Bar{\phi}$, the inhomogeneity rapidly decays, resulting in the disappearance of the excluded volume effects. It becomes apparent that the failure of the hypothetical argument stated in the paragraph written in italics is due to the neglect of the inhomogeneity term. The same conclusion can be drawn more confidently by the use of the legitimate approximation (\ref{Nenk-11}$'$) (see the next section and Fig. \ref{Nenkai-10}).

The present discussion reveals another aspect that ``the asymptotic limit of $n/V\rightarrow 0$ is a different physical concept from an isolated polymer system (a single polymer in a solution).''\\[-3mm]

Then let us proceed to the main discussion of this paper.

\vspace*{0mm}
\section{The Improved Theory and Test by the Neutron Scattering Experiments}\label{BBP}
\begin{figure}[H]
\begin{center}
\begin{minipage}[t]{0.46\textwidth}
\begin{center}
\includegraphics[width=7.8cm]{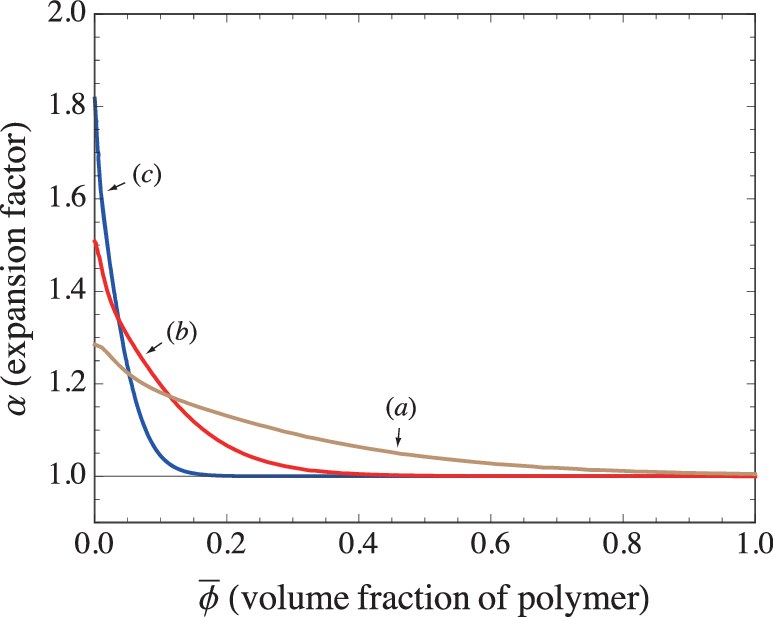}
\end{center}
\vspace{-2mm}
\caption{Molecular weight dependence of $\alpha$ calculated by Eq. (\ref{Nenk-10}) for PSt$-\text{CS}_{2}$: (a) $\text{M}=10^{4}$, (b) $\text{M}=10^{5}$, and (c) $\text{M}=10^{6}$.}\label{Nenkai-3}
\end{minipage}
\hspace{10mm}
\begin{minipage}[t]{0.46\textwidth}
\begin{center}
\includegraphics[width=7.8cm]{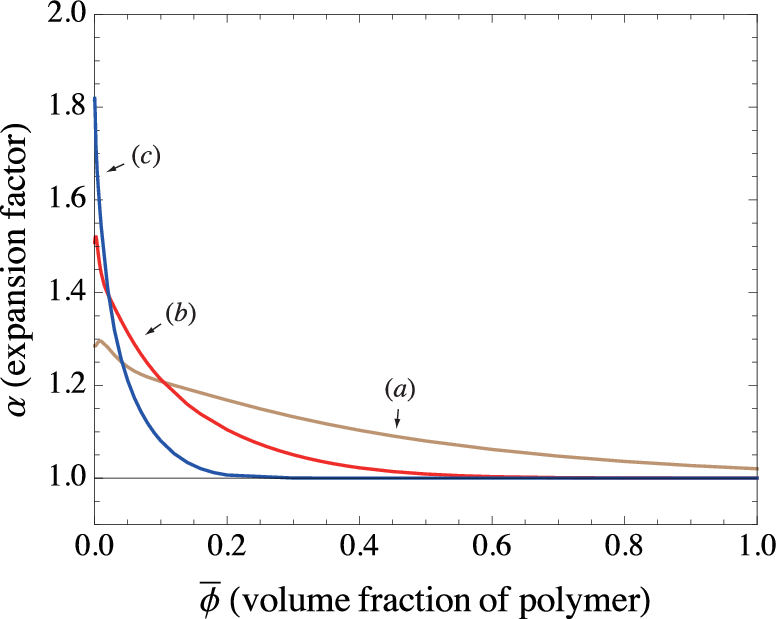}
\end{center}
\vspace{-2mm}
\caption{Molecular weight dependence of $\alpha$ calculated by Eq. (\ref{Nenk-11}$'$) for PSt$-\text{CS}_{2}$: (a) $\text{M}=10^{4}$, (b) $\text{M}=10^{5}$, and (c) $\text{M}=10^{6}$.}\label{Nenkai-4}
\end{minipage}
\end{center}
\vspace*{0mm}
\end{figure}

\begin{figure}[H]
\begin{center}
\begin{minipage}[t]{0.46\textwidth}
\begin{center}
\includegraphics[width=7.8cm]{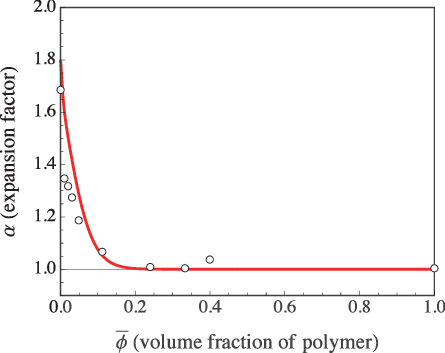}
\end{center}
\vspace{-2mm}
\caption{Concentration dependence of $\alpha$ calculated by Eq. (\ref{Nenk-10}) for PMMA$-$CHCl$_{3}$ ($N=5900$). The solid line is the theoretical line for $\chi=0.3$; open circles ($\circ$): observed points by Cheng, Graessley, and Melnichenko\cite{Graessley}.}\label{Nenkai-5}
\end{minipage}
\hspace{10mm}
\begin{minipage}[t]{0.46\textwidth}
\begin{center}
\includegraphics[width=7.8cm]{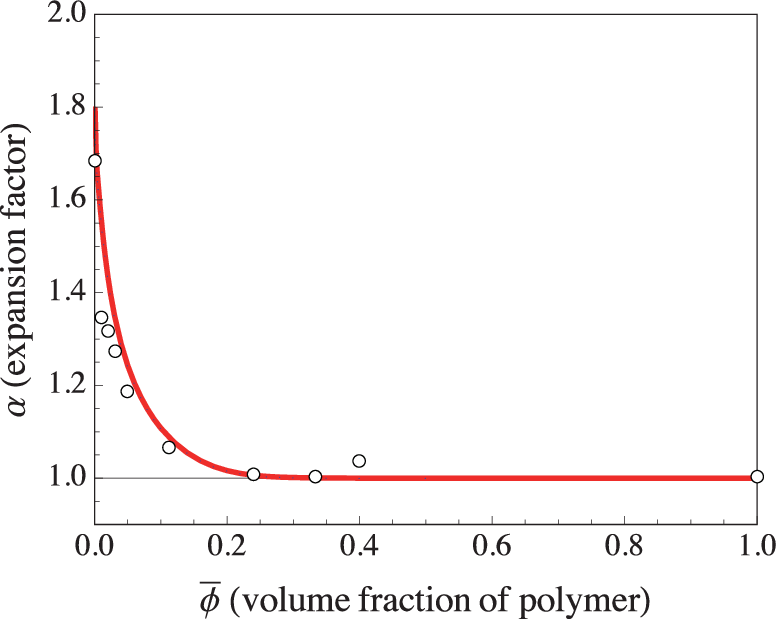}
\end{center}
\vspace{-2mm}
\caption{Concentration dependence of $\alpha$ calculated by Eq. (\ref{Nenk-11}$'$) for PMMA$-$CHCl$_{3}$ ($N=5900$). The solid line is the theoretical line for $\chi=0.3$; open circles ($\circ$): observed points by Cheng, Graessley, and Melnichenko\cite{Graessley}.}\label{Nenkai-6}
\end{minipage}
\end{center}
\vspace*{-4mm}
\end{figure}

Let us examine the essence of the equation:
\begin{equation}
\alpha-1/\alpha =-\frac{1}{3V_{1}}\frac{\partial}{\partial\alpha}\iiint\left\{\left(1/2-\chi\right)\Hat{\mathscr{J}}^{2}+\frac{1}{6}\Hat{\mathscr{J}}^{3}+\cdots\right\}dxdydz\tag{\ref{Nenk-11}$'$}
\end{equation}
derived by the legitimate approximation (\ref{Nenk-4}) and compare with the original equation (\ref{Nenk-10}) derived by Eq. (\ref{Nenk-2}). Eq. (\ref{Nenk-11}$'$) can be solved numerically according to the lattice model. We again consider the simple cubic lattice with the unit lengths $p\times p\times p$, on which each polymer is put on the site. The inhomogeneity terms are of the form:
\begin{equation}
\Hat{J}_{\alpha}^{k}=\iiint_{-p/4}^{p/4}\HG\hspace{0.3mm}^{\,k}\,dxdydz-\iiint_{p/4}^{3p/4}\HG\hspace{0.3mm}^{\,k}\,dxdydz\label{Nenk-16}
\end{equation}
($k=2, 3, \cdots$). The equations (\ref{Nenk-10}) and (\ref{Nenk-11}$'$) yield different answers, as a matter of course. Remarkably, however, if we restrict our discussion to linear polymer systems, the difference does not arise very pronouncedly. The typical examples are shown for the polystyrene(PSt)$-\text{CS}_{2}$ system in Figs. \ref{Nenkai-3} and \ref{Nenkai-4}, and for the poly(methyl methacrylate)(PMMA)$-$CHCl$_{3}$ system in Figs. \ref{Nenkai-5} and \ref{Nenkai-6}. Here, Figs. \ref{Nenkai-3} and \ref{Nenkai-5} show the simulation results according to Eq. (\ref{Nenk-10}), while Figs. \ref{Nenkai-4} and \ref{Nenkai-6} the results calculated according to Eq. (\ref{Nenk-11}$'$). Comparing Figs. \ref{Nenkai-5} and \ref{Nenkai-6}, it is seen that Eq. (\ref{Nenk-11}$'$) faithfully reproduces the experimental points by Cheng, Graessley, and Melnichenko\cite{Graessley}, implying that Eq. (\ref{Nenk-11}$'$) is a more proper formulation than Eq. (\ref{Nenk-10}). Nonetheless, putting aside some deviation from the experimental points observed in Fig. \ref{Nenkai-5}, Eq. (\ref{Nenk-10}) extracts the essential features of the excluded volume phenomena within experimental errors, i.e., it reproduces well the swollen-to-unperturbed coil transition point, the disappearance point of the excluded volume effects\cite{Daoud, Westermann, Graessley}.

\begin{table}[H]
\caption{Basic parameters of poly(styrene) and poly(methyl methacrylate) solutions.}\label{PSt-table1}
\begin{center}
\vspace*{-4mm}
\begin{tabular}{l l c r}\hline\\[-2.5mm]
& \hspace{10mm}parameters & notations & values \,\,\,\,\\[2mm]
\hline\\[-1.5mm]
poly(styrene) (PSt) & volume of a solvent (CS$_{2}$) & $V_{1}$ & \hspace{5mm}100 \text{\AA}$^{3}$\\[1.5mm]
& volume of a segment (C$_{8}$H$_{8}$) & $V_{2}$ & \hspace{5mm}165 \text{\AA}$^{3}$\\[1.5mm]
& Flory characteristic ratio & $C_{\textit{\textsf{\hspace{-0.3mm}F}}}$\index{Flory characteristic ratio $C_{\textit{\textsf{\hspace{-0.3mm}F}}}$} & \hspace{5mm}10 \,\,\,\,\,\,\,\\[1.5mm]
& mean bond length & $\bar{l}$ & \hspace{5mm}1.55 \text{\AA}\,\,\,\\[1.5mm]
& enthalpy parameter (25$^{\,\circ}$C) & $\chi$ & \hspace{5mm}0.4 \,\,\,\,\,\,\,\\[2mm]
\hline
\hline\\[-2.5mm]
poly(methyl methacrylate) & volume of a solvent (CHCl$_{3}$) & $V_{1}$ & \hspace{5mm}$134\,\text{\AA}^{3}$\\[1.5mm]
& volume of a segment (C$_{5}$O$_{2}$H$_{8}$) & $V_{2}$ & \hspace{5mm}$140\,\text{\AA}^{3}$\\[1.5mm]
& Flory characteristic ratio & $C_{\textit{\textsf{\hspace{-0.3mm}F}}}$\index{Flory characteristic ratio $C_{\textit{\textsf{\hspace{-0.3mm}F}}}$} & \hspace{5mm}9.2\,\,\,\,\,\,\,\,\\[1.5mm]
& mean bond length & $\bar{l}$ & $\hspace{5mm}1.56\text{\AA}$\,\,\,\\[1.5mm]
& enthalpy parameter (25$^{\,\circ}$C) & $\chi$ & \hspace{5mm}$0.3$\,\,\,\,\,\,\,\,\\[2mm]
\hline\\[-6mm]
\end{tabular}\\[6mm]
\end{center}
\end{table}

It is noteworthy that the quantity $p$ introduced in Eq. (\ref{Nenk-1}), despite its improper formulation, has made the unexpected success. The reason becomes immediately clear by taking notice that the inhomogeneity term $[\left(\tfrac{2\beta}{\pi}\right)^{3/2}\iiint\left(\G\hspace{0.3mm}_{hill}^{\,2}-\G\hspace{0.3mm}_{valley}^{\,2}\right)dxdydz]$ in Eq. (\ref{Nenk-10}) decreases very steeply with increasing $N$ and $\bar{\phi}$ as is seen from Figs. \ref{Nenkai-1} and \ref{Nenkai-2}, which conveniently obscures the difference in the exponential terms between $\G$ (without $\alpha$) and $\HG$ (with $\alpha$). In compensation for the use of the improper probability distribution function $p$, Eq. (\ref{Nenk-10}) has gained some merits if we restrict our discussion to linear polymer systems. Foremost is that the formulation (\ref{Nenk-10}) is a closed solution, and for this reason, the physical interpretation of the excluded volume phenomena in concentrated solutions, together with the applicability limit of the equation (\ref{Nenk-10}), is easy to understand. This is the reason why we have occasionally made use of Eq. (\ref{Nenk-10}), as we have done in the discussion of Section \ref{Theo}\cite{Kazumi};  ``the pros outweigh the cons.''

\vspace*{3mm}

\section{Application to the Extended Comb Polymer with $\mb{\nu_{0}=1/4}$}\label{AECP}
The situation changes markedly when we apply the equations (\ref{Nenk-10}) and (\ref{Nenk-11}$'$) to branched polymer systems. The difference between the inhomogeneity functions, $\G$ (without $\alpha$) and $\HG$ (with $\alpha$), becomes pronounced, which was hidden for linear polymers behind the conformational characteristics.

\begin{shaded}
\vspace*{-3mm}
\subsection*{Difference in Inhomogeneity terms that result from Eqs. (\ref{Nenk-1}) and (\ref{Nenk-3})}
$\textbf{p}$:
\begin{align}
J_{\alpha}&=\iiint_{hill}\G^{\,2}\,dxdydz-\iiint_{valley}\G^{\,2}\,dxdydz\notag\\
&\hspace{2cm}\G=\sum_{\{a, b, c\}}\exp\left\{-\beta\left[(x-a)^{2}+(y-b)^{2}+(z-c)^{2}\right]\right\}\tag{\ref{Nenk-15}}
\end{align}
$\Hat{\textbf{p}}$:
\begin{align}
\hspace{3.1cm}\Hat{J}_{\alpha}^{\,k}&=\iiint_{hill}\HG^{\,k}\,dxdydz-\iiint_{valley}\HG^{\,k}\,dxdydz\notag\\
&\hspace{2cm}\HG(x, y, z)=\sum_{\{a, b, c\}}\exp\left\{-\frac{\beta}{\alpha^{2}}\left[(x-a)^{2}+(y-b)^{2}+(z-c)^{2}\right]\right\}\hspace{1cm}\tag{\ref{Nenk-16}}
\end{align}
\end{shaded}
As mentioned above, the pronounced reduction of the segment density in linear polymers, with increasing $N$ and $\Bar{\phi}$, has obscured the difference between $\G$ (without $\alpha$) and $\HG$ (with $\alpha$). The failure of the approximation, $p$, in the application to branched polymer systems can be ascribed to the following points: The excluded volume effects are the direct consequence of the inhomogeneity in the solution; the inhomogeneity tends to diminish with the interpenetration of segments, and the interpenetration will be enhanced both by (i)  polymer concentration and (ii) the expansion of a polymer. It is seen from Eq. (\ref{Nenk-15}) that $J_{\alpha}$ lacks this second factor, the effect of the volume expansion. Thus, $J_{\alpha}$ is incomplete while $\Hat{J}_{\alpha}^{\,k}$ is exact. For branched polymer systems, Eq. (\ref{Nenk-10}) fails, and we must use Eq. (\ref{Nenk-11}$'$).

We simulate Eq. (\ref{Nenk-11}$'$), as functions of $\Bar{\phi}$ and $N$, for the extended comb polymer (equivalent to the $z=2$ polymer), which has the same length of side chains as the backbone and has the mean square of the radius of the form:

\begin{equation}
\left\langle s_{N}^{2}\right\rangle_{0}=\frac{1}{6}\frac{(N-1)(4\sqrt{N}-3)}{N}\,l^{2}\label{Nenk-17}
\end{equation}

\begin{figure}[h]
\begin{center}
\includegraphics[width=15cm]{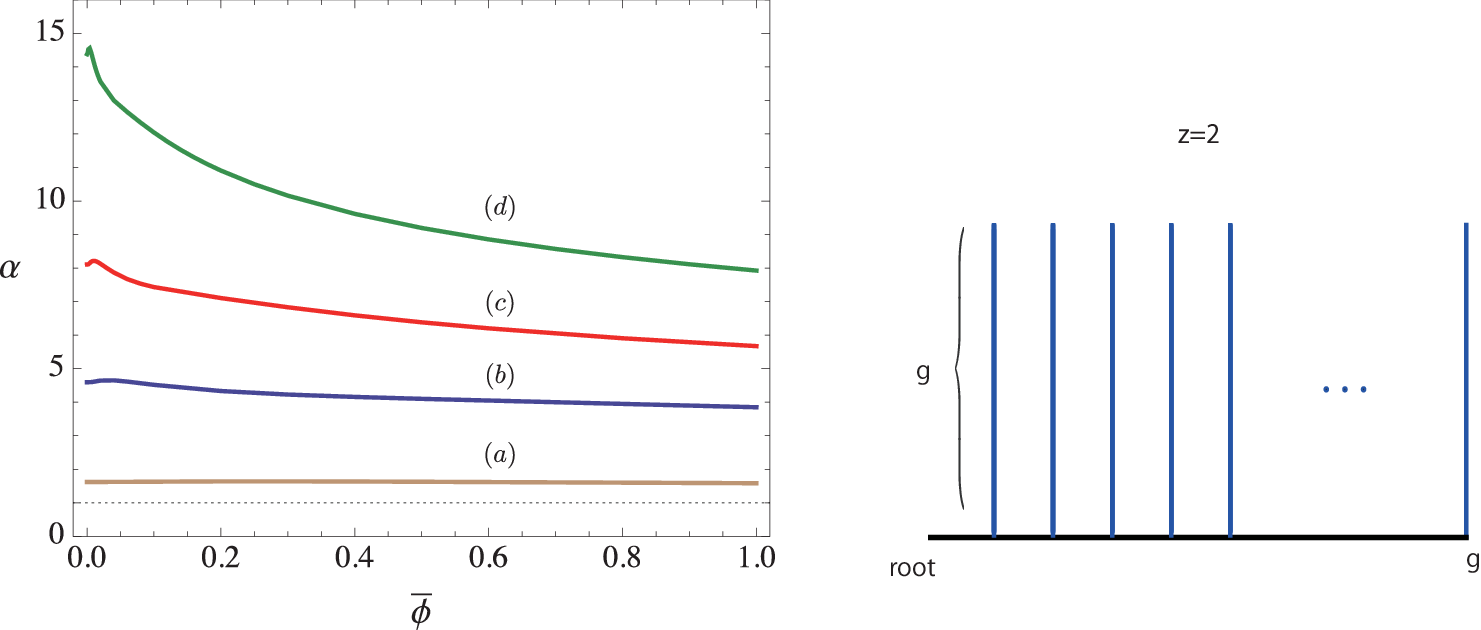}
\caption{The average volume fraction $\bar{\phi}$ dependence of the expansion factor, $\alpha$, for the $z=2$ polymer ($\nu_{0}=1/4$) with (a) $N=10^{2}$, (b) $N=10^{4}$, (c) $N=10^{5}$, and (d) $N=10^{6}$, ($d=3$). Calculated numerically according to Eq. (\ref{Nenk-11}$'$) with the help of Eq. (\ref{Nenk-16}).}\label{Nenkai-7}
\end{center}
\vspace*{-3mm}
\end{figure}

\noindent Let a monomer unit of this polymer be composed of the methylene unit. Then, the molecular mass is $M=14\,N+2$. Employed parameters are $V_{1}=569\,\text{\AA}$ ($n$-nonadecane), $V_{2}=25\,\text{\AA}$ ($-\text{CH}_{2}-$: a segment), $\chi=0.2$, and $l=1.54\,\text{\AA}$, mimicking the PE$-$$n$-nonadecane system. As one can see from Fig. \ref{Nenkai-7}, for the polymer having $N=10^{2}$, the dimensions are almost invariable over the whole concentration range, indicating the absence of segment interpenetration among different polymers. On the other hand, for larger polymers, the dimensions decrease with $\Bar{\phi}$ as expected. Noteworthy is the fact that those branched polymers show extremely large $\alpha$'s over the whole ranges of $\Bar{\phi}$ and $N$, in striking contrast to the case of linear polymers. More important is the fact that the expansion factor is greatly enhanced with increasing $N$ in all the concentration ranges, contrary to the linear case. Such a large $N$ dependence of $\alpha$ is inevitable for the branched polymers not to collapse below the critical packing density, because it simply approaches to the critical value, 1/3.

\begin{figure}[H]
\begin{center}
\begin{minipage}[t]{0.46\textwidth}
\begin{center}
\includegraphics[width=8cm]{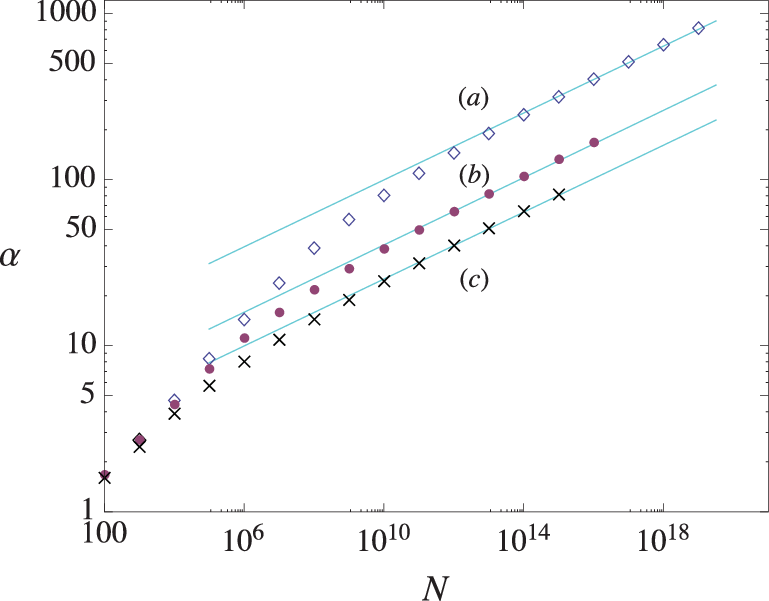}
\end{center}
\vspace{-2mm}
\caption{Expansion factor as against $N$ for the $z=2$ polymer in $n$-nonadecane ($V_{1}=569\text{\AA}^{3}$, $V_{2}=25\text{\AA}^{3}$, $\bar{l}=1.54\text{\AA}$, $\chi=0.2$, $d=3$). The plot points were calculated according to Eq. (\ref{Nenk-11}$'$) for (a) $\bar{\phi}=0.01$, (b) $\bar{\phi}=0.2$, (c) $\bar{\phi}=1\, (melt)$.}\label{Nenkai-8}
\end{minipage}
\hspace{10mm}
\begin{minipage}[t]{0.46\textwidth}
\begin{center}
\includegraphics[width=8.2cm]{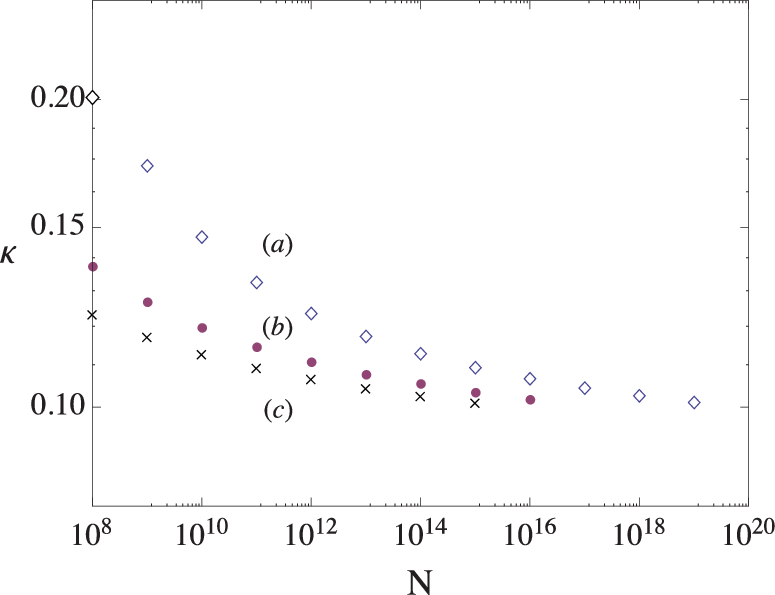}
\end{center}
\vspace{-2mm}
\caption{The gradient, $\kappa=\log\alpha/\log N$, as against $N$ for the $z=2$ polymer in $n$-nonadecane ($V_{1}=569\text{\AA}^{3}$, $V_{2}=25\text{\AA}^{3}$, $\bar{l}=1.54\text{\AA}$, $\chi=0.2$, $d=3$). The plot points were calculated according to the data in Fig. \ref{Nenkai-8} for (a) $\bar{\phi}=0.01$, (b) $\bar{\phi}=0.2$, (c) $\bar{\phi}=1\, (melt)$.}\label{Nenkai-9}
\end{minipage}
\end{center}
\vspace*{-4mm}
\end{figure}

\begin{figure}[H]
\begin{center}
\includegraphics[width=7.8cm]{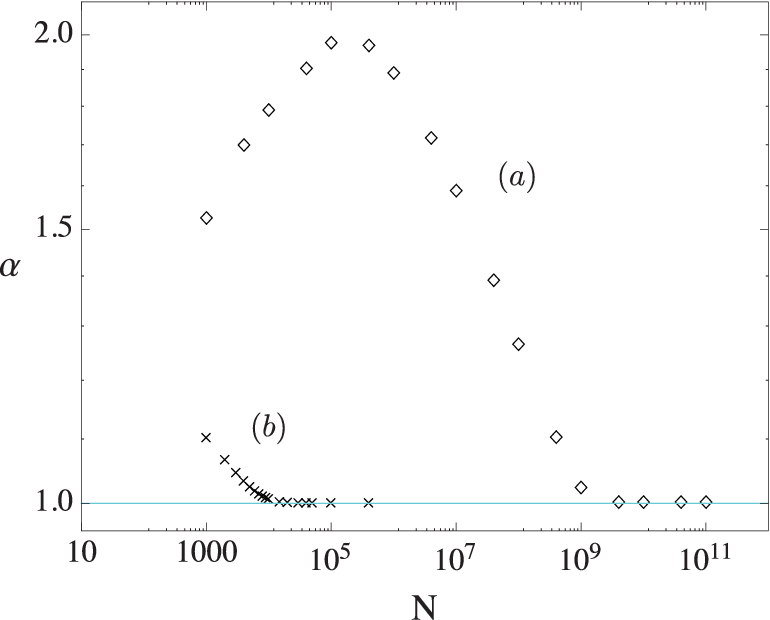}
\vspace{-2mm}
\caption{Expansion factor as against $N$ for the PSt$-$CS$_{2}$ solution ($V_{1}=100\text{\AA}^{3}$, $V_{2}=165\text{\AA}^{3}$, $\bar{l}=1.55\text{\AA}$, $\chi=0.4$, $d=3$). The plot points were calculated according to Eq. (\ref{Nenk-11}$'$) for (a) $\bar{\phi}=0.001$, and (b) $\bar{\phi}=0.2$.}\label{Nenkai-10}
\end{center}
\vspace*{-4mm}
\end{figure}

Next, with the help of Eq. (\ref{Nenk-11}$'$), we calculated the gradient, $\kappa=\log\alpha/\log N$, for the $z=2$ polymer$-$$n$-nonadecane system (Figs. \ref{Nenkai-8} and \ref{Nenkai-9}) and compared with the PSt$-$CS$_{2}$ system (Fig. \ref{Nenkai-10}). As one can see, whereas the linear system rapidly approaches $\kappa=0$ as $N\rightarrow\infty$, the branched system shows $\kappa\simeq 0.1$ for large $N'$s, irrespective of the polymer volume fraction $\Bar{\phi}$ (Fig. \ref{Nenkai-9}); however, the gradient shown in Fig. \ref{Nenkai-9} is not in the asymptotic limit at all at this stage, but is still decreasing gradually with $N$ and appears to descend toward the critical packing density $\kappa=1/12$ (or equivalently $\nu=1/3$). So we expect the scaling relation: $\alpha\propto N^{1/12}$ for $N\rightarrow\infty$  for the branched polymer system of all the finite concentration ranges, $0<\Bar{\phi}\le 1$. The results are consistent with the previous observations for the randomly branched polymer\cite{Kazumi}. In summary, the expansion factor obeys the asymptotic relations of $N\rightarrow\infty$:
\begin{equation}
\alpha\propto
\begin{cases}
\hspace{1mm}N^{0} & \hspace{3mm}\mbox{for linear polymers (with}\hspace{2mm}\displaystyle\nu_{0}=1/2)\\[3mm]
\hspace{1mm}N^{1/12}& \hspace{3mm}\mbox{for branched polymers (with}\hspace{2mm}\displaystyle\nu_{0}=1/4)
\end{cases}\label{Nenk-18}
\end{equation}
for all the finite concentration ranges $0<\bar{\phi}\le1$.

\section{Concluding Remarks}
In conjunction with the previous findings, it may be concluded that the linear polymer approaches the ideal chain as $N\rightarrow\infty$ for an arbitrary concentration in the interval, $0<\bar{\phi}\le1$, while the branched polymer (with $\nu_{0}=1/4$) seems to approach $\alpha\propto N^{1/12}$, namely $\nu=1/3$ for $N\rightarrow\infty$. The results are summarized in Table \ref{AlD-Table2}.
\vspace*{0mm}
 \begin{table}[h]
 \centering
  \begin{threeparttable}
    \caption{Values of the exponents, $\nu$, for the branched polymers with $\nu_{0}=1/4$, and those for linear polymers in good solvents ($\chi'>0$\,\, or $T>\Theta$). $\langle s_{N}^{2}\rangle\propto N^{2\nu}$ ($d=3$).\vspace*{-2mm}}\label{AlD-Table2}
  \begin{tabular}{l l c c}
\hline\\[-1.5mm]
\hspace{3mm} systems & \hspace{3mm} exponents $\nu$ &  & \hspace{5mm} concentrations \,\,\,\,\\[2mm]
\hline\\[-1.5mm]
\hspace{3mm} linear polymer ($\nu_{0}=\frac{1}{2}$) & \hspace{5mm} $\nu_{good}=\frac{3}{5}$  &  & \hspace{0mm} dilution limit\\[2.5mm]
& \hspace{5mm} $\nu_{good}=\frac{1}{2}$  & & \hspace{1mm} $0<\bar{\phi}\le1$\\[1.5mm]
\hline
\hline\\[-1.5mm]
\hspace{3mm} branched polymer ($\nu_{0}=\frac{1}{4}$) & \hspace{5mm} $\nu_{good}=\frac{1}{2}$  &  & \hspace{0mm} dilution limit\\[2.5mm]
& \hspace{5mm} $\nu_{good}=\frac{1}{3}$  & & \hspace{1mm} $0<\bar{\phi}\le1$\\[2mm]
\hline\\[-6mm]
   \end{tabular}
      \vspace*{2mm}
   \begin{tablenotes}
     \item $\chi'=1/2-\chi$;\,\, $\bar{\phi}$ is the average volume fraction of polymer in the whole system.
   \end{tablenotes}
  \end{threeparttable}
  \vspace*{4mm}
\end{table}
\begin{description}

\item[1.]  The approximate expression $p$ for the expanded polymer works well for the polymers having $\nu_{0}=1/2$, but not for the branched polymers having $\nu_{0}=1/4$, whereas $\Hat{p}$ works well irrespective of $\nu_{0}$. As expected, Eq. (\ref{Nenk-11}$'$) (theorized by $\Hat{p}$) reproduces the experimental observations (Fig. \ref{Nenkai-6})\cite{Daoud, Westermann, Graessley} more faithfully than Eq. (\ref{Nenk-10}) (theorized by $p$). Thus, Eq. (\ref{Nenk-11}$'$) is a legitimate formulation for the expanded polymers.
\item[2.]  The size exponent has a critical nature. For linear polymers in good solvents, it varies abruptly from $3/5$ in the isolated system to $1/2$ in the finite concentration ($0<\bar{\phi}\le1$), whereas for branched polymers with $1/4$, it varies from $1/2$ in the isolated system to $1/3$ in the finite concentration, discontinuously. There are no intermediate values between them.
\end{description}
The above conclusions reconfirm the previous works\cite{Kazumi}.

\vspace{5mm}

\end{document}